\documentstyle[prb,aps,epsf]{revtex}

\begin{document}
\title{Magnetoresistance in quasi-one-dimensional metals due
to Fermi surface cold spots}
\draft
\author{Perez Moses and Ross H. McKenzie\cite{email} }
\address{School of Physics, University of New
South Wales, Sydney 2052, Australia}
\date{\today}
\maketitle
\widetext
\begin{abstract}
In a number of quasi-one-dimensional organic metals
the dependence 
of the magnetoresistance on the direction of the magnetic field 
is quite different from the predictions of
Boltzmann transport theory for a Fermi liquid
with a scattering rate that is independent of momentum.
We consider a model in which there are
large variations in the scattering rate over
the Fermi surface.
The model is the quasi-one-dimensional version
of the ``cold spots'' model introduced
by Ioffe and Millis to explain anomalous transport properties
of the metallic phase of the cuprate superconductors.
The dependence of the resistance,
in the most and least conducting directions,
on the direction and magnitude of the magnetic field
are calculated.
The calculated magnetoresistance has a number of
properties
that are quite distinct from conventional transport theory
such as magic angle effects and  a significant magnetoresistance when
the field and current are both in the least conducting direction.
However, the model cannot give a complete
description of the unusual properties of
(TMTSF)$_2$PF$_6$ at pressures of 8-11 kbar.
\end{abstract}

\section{Introduction}

Many of the electronic transport properties
of strongly correlated metals such as
cuprate superconductors,\cite{anderson,liang}
heavy fermions,\cite{degiorgi}
and organic superconductors\cite{wosnitza,mck,jerome} are significantly
different from elemental metals. The transport
properties of the latter
are adequately described by Boltzmann transport
theory, which is based on a Fermi liquid
picture, in which there is one-to-one correspondence
between the elementary excitations and those
of a non-interacting Fermi gas.\cite{AM}
An important and controversial question is whether
to describe strongly correlated metals one must
completely abandon Fermi liquid theory or
whether one can just make modest modifications
to Fermi liquid theory, such as allowing
the scattering rate
to vary significantly over different parts of the Fermi surface.
An example of the former point of view
for the cuprates is that of
Anderson\cite{anderson} and of the
latter is that of Pines\cite{pines} or Ioffe and Millis.\cite{ioffe:cs}
For heavy fermions near a quantum critical point,\cite{mathur}
the former point of view has been advocated by
Coleman\cite{coleman} and Smith and Si,\cite{si}
and the latter by Rosch.\cite{rosch}
The only way to resolve this issue is to
perform calculations for specific
models in order to produce predictions
that can be used to falsify that model.

The theoretical description of the magnetoresistance
of the metallic phase of the Bechgaard salts, (TMTSF)$_2$X [where TMTSF
is the tetramethyl-tetraselenafulvane
molecule and X is an anion] represents a considerable
challenge. The experimental data is briefly summarised
below. Strong, Clarke, and Anderson\cite{sca:cs} and
Zheleznyak and Yakovenko\cite{yak1} have argued that
the data imply a non-Fermi liquid description
whereas many others\cite{mak,osa2,blundell,lebbak,chaikin:cs}
have tried to explain the data within a Fermi liquid description.
None of these theories gives a complete description
of the experimental data.
The purpose of this paper is to calculate
the properties of the magnetoresistance
within a ``cold spots'' model
(where the scattering rate varies over the Fermi surface).
This model is the quasi-one-dimensional version of
a model originally proposed for the cuprates
by Ioffe and Millis.\cite{ioffe:cs}
The model has the distinct advantage that it is analytically
tractable, allowing the calculation of a
wide range of properties of the magnetoresistance
that can be compared to experimental results.

We now briefly summarise the observed properties of
the magnetoresistance of (TMTSF)$_2$X that
cannot be explained with Boltzmann
transport theory with a simple dispersion relation
and a scattering rate that
is constant over the Fermi surface.
The most puzzling data is that of (TMTSF)$_2$PF$_6$
at pressures of about 10 kbar.\cite{clarke}
We also note that the
magnetoresistance of the quasi-two-dimensional metal
$\alpha$-(BEDT-TTF)$_2$MHg(SCN)$_4$ [M = K,Rb,Tl] also
exhibits unusual temperature and angular dependence.\cite{mcken,qualls} 

1. {\it The magic angle effect}. When the magnetic
field is rotated in the plane perpendicular to
the most conducting direction (i.e., in the $b-c$ plane)
one observes dips in the resistance versus angle curve
at angles, (where $\theta$ is the angle between the
field direction and the c-axis) such that
$\tan \theta = n b/c$ where $b$ and $c$ are lattice
constants and $n=1,2,...$. The features
at $n=1$ and 2 are most prominent.

2. {\it Angular dependence.}
The simplest Boltzmann transport models predict
no magnetoresistance when the magnetic field
and current are parallel and
the magnetoresistance is a maximum
when the field and current are perpendicular.
This is observed in (TMTSF)$_2$ClO$_4$ at ambient and 6 kbar
pressure\cite{dan:cs,chas:cs}. However, the
opposite is observed in (TMTSF)$_2$PF$_6$ at 10 kbar:
the magnetoresistance is much larger
when the field and current are parallel
than when they are perpendicular.\cite{dan:cs}.
Specifically, the background magnetoresistance
(i.e., after the magic angle effect is subtracted out)
only depends on the component of the field
perpendicular to the layers.
Furthermore, for moderate fields the
resistivity in the most conducting direction,
$\rho_{xx} \sim (B \cos \theta)^{0.5}$ and
the resistivity in the
least conducting direction
$\rho_{zz} \sim (B \cos \theta)^{1.3}$.
Simple Fermi liquid theory would generally not
produce such a non-integer exponent.
Note, that this means that there is no magnetoresistance
for fields parallel to the b-axis.

3. {\it Kohler's rule.}
In a conventional metal with a single scattering rate
this provides a simple
way to relate the field and temperature
dependence of the resistance.
In (TMTSF)$_2$ClO$_4$ at ambient pressure\cite{ong}
and at 6kbar\cite{chas:cs} this is satisfied.
However, in (TMTSF)$_2$PF$_6$ at 10 kbar
there are large violations.

In order to explain the magic angle effect
Chaikin first proposed a ``hot spots'' model,
where the scattering rate is significantly larger
than elsewhere on the Fermi surface.\cite{chaikin:cs}
Zheleznyak and Yakovenko did find that
the scattering rate due to electron-electron scattering
exhibited hot spots. However, these were not
of sufficient strength to produce a large
magnetoresistance or the magic angle effect.\cite{zhel:cs}

In order to explain the anomalous transport
properties of cuprates several authors have considered
the effects of both hot spots\cite{carr:cs,sto:cs,hlub:cs}
and cold spots\cite{ioffe:cs,yak,marel}.
Ioffe and Millis\cite{ioffe:cs} considered
a cold spot model where the scattering rate variation had the
same symmetry (d-wave) as the superconducting order parameter, i.e.,
the cold spots are associated with nodes in the energy gap
(or pseudogap)
which exists in the superconducting phase. Although it
is not clear what specific
microscopic mechanism produces the cold spots,
Ioffe and Millis suggest that they
might arise from strong superconducting pairing fluctuations.
The model provides a simple explanation of
photoemission experiments which show
that in the
cuprates the electron spectral function varies significantly
over the Fermi surface. Along the zone diagonals
the spectral function has a well defined quasi-particle peak, suggesting
weak scattering; in other regions the spectral function is broad,
suggesting strong scattering.
Using this simple model and a Boltzmann equation
analysis, Ioffe and Millis reproduced quantitatively the
frequency and temperature
dependence of the observed dc and ac, longitudinal, and Hall conductivities
in the cuprates.
However, the calculated magnetoresistance is much larger in magnitude
and has a stronger temperature dependence than is observed.

In this paper we investigate to what
extent such a cold spot model can explain the anomalous
magnetoresistance in the quasi-one-dimensional metals, (TMTSF)$_2$X.
We find that the calculated magnetoresistance
does have a number of unusual features that are consistent with experiment.
(i) When the magnetic field and current are parallel to
the least-conducting direction, there is a large
positive magnetoresistance.
This increases with the strength of the cold spots.
(ii) When the magnetic field is rotated in
the $b-c$ plane the resistivity in the most-conducting direction
has an angular dependence qualitatively
similar to the background magnetoresistance
of (TMTSF)$_2$PF$_6$ at 10 kbar.
The resistance is largest when the field
is in the least-conducting direction. Furthermore, it
only depends on the component of the field parallel
to the least conducting direction.
(iii) Magic angle effects do occur in the interlayer resistance.

However, there are a number of properties that
are inconsistent with experiment.
(a) The magnetoresistance saturates with
increasing field when the magnetic field and current are parallel to
the least-conducting direction.
(b) No magic angle effects occur in the
resistivity in the most-conducting direction.
(c) For reasonable strengths of the magnetic
field the size of the features in the interlayer
resistance at the magic angles
are much smaller than is observed.
Further, peaks rather than dips are predicted
at the odd-integer magic angles.
(d) When the magnetic field is parallel to
the b-axis the interlayer magnetoresistance
increases quadratically with field,
whereas in (TMTSF)$_2$PF$_6$ at 10 kbar,
it saturates with increasing field.
Table \ref{table1:cs} 
gives a brief summary of the successes and
failures of the cold spot model.

The outline of the paper is as follows.
In Section \ref{deriv-cond}
the Boltzmann equation is solved
in the relaxation time approximation
for the general case of a scattering rate
that varies over the Fermi surface.
We introduce the specific model for the momentum dependence
of the scattering rate that we use.
It is shown that in zero field the resistivity
is proportional to the inverse of the average
of the scattering time over the Fermi surface.
In the high field limit the resistivity
is proportional to the average of the scattering
rate over the Fermi surface.
We then show by the use of the Cauchy-Schwarz inequality
that the resistance at high fields will always be larger
than the resistance at zero field.
In Section \ref{cond-zero} the interlayer conductivity in
zero field is explicitly evaluated and we consider
different models for its temperature dependence.
In Section \ref{cond-all} the interlayer conductivity
is calculated for various directions of the magnetic field.
Section \ref{condx-axis} 
contains a similar calculation for the conductivity
in the most conducting direction.

\section{BOLTZMANN TRANSPORT THEORY WITH A MOMENTUM-DEPENDENT
SCATTERING RATE}
\label{deriv-cond}
\subsection{Derivation of the conductivity}
\label{introcc}
If the scattering rate does not vary over the Fermi surface
then the Boltzmann equation can be solved in the relaxation
time approximation to yield Chamber's formula for the
conductivity in the presence of a magnetic field.\cite{AM} We now
consider how this is modified in the presence of a
scattering rate that varies over the Fermi surface. Following
Ashcroft and Mermin (p.246ff)\cite{AM}, let $g(\vec{r},\vec{k},t)$ be the
non-equilibrium distribution function which describes the probability of
finding the electron at $\vec{r}$ with momentum $\vec{k}$ at
time $t$. $P(t,t^{'})$ denotes the fraction of electrons that
are not scattered between times $t$ and $t^{'}$ and satisfies
the differential equation
\begin{equation}
{\partial \over \partial t^{'}} P(t,t^{'}) = {P(t,t^{'}) \over \tau(t^{'})} \ 
\end{equation}
where $\tau(t) = \tau(\vec{k}(t))$.
Integrating this gives
\begin{equation}
P(t,t^{'})= \exp\left( -\int_{t^{'}}^{t} {du \over \tau(u)} \right)\ .
\end{equation}
The non-equilibrium distribution function can then be written as
\begin{equation}
g({\vec{r},\vec{k},t}) = f -
{\partial f \over \partial E }\int_{-\infty}^{t} dt^{'} \vec{E}\cdot
\vec{v} P(t,t^{'}) \ ,
\end{equation}
where $f(E)$ is the Fermi function and equals the equilibrium distribution
and $\vec{E}$ is the electric field. The conductivity then reduces to
\begin{equation}
\sigma_{ij} = {e^2 \over 4 \pi^3}
\int { v_i(\vec k) \bar{v}_j(\vec k)}
\left(-{\partial f(E) \over \partial E}\right) d^3\vec{k}\ ,
\label{condcs}
\end{equation}
where $\bar{v}_j(\vec k)$ is 
\begin{equation}
\label{tauont}
\bar{v}_j(\vec k)= \int_{-\infty}^{0}
\exp\left[ -\int_{0}^{t} {du \over \tau(\vec{k}(u))}\right]
v_{j}(\vec{k}(t)) dt
\end{equation}
and the wave vector $\vec{k}(t)$ satisfies the semi-classical
equation of motion
\begin{equation}
{d\vec{k} \over dt} = -{e \over h^2} \vec{\nabla}_k
\epsilon(\vec{k})\times \vec{B} \ .
\end{equation}

In a quasi-one dimensional metal the simplest possible
dispersion relation is
\begin{equation}
\label{disper}
\epsilon (\overrightarrow{k})=\hbar v_{F}(|k_{x}|-k_{F})-2t_{b}
\cos (bk_{y})-2t_{c}\cos (ck_{z}) \ ,
\end{equation}
where $v_F$ is the Fermi velocity, $k_F$ is the Fermi wave vector,
and $t_b$ and $t_c$ are the electron hopping integrals perpendicular
to the chains.
For the dispersion (\ref{disper}), the interlayer conductivity given
by Eq.~(\ref{condcs}) reduces to
\begin{equation}
\sigma_{zz} = {e^2 \over 4 \pi^3 \hbar v_F}
\int_{-\pi/c}^{\pi/c} dk_z(0)
\int_{0}^{2\pi/b} dk_y(0) { v_z(\vec k) \bar{v}_z(\vec k)} \
\end{equation}
assuming that the temperature is sufficiently low that
the derivative of the Fermi function can be
replaced by a delta function at the Fermi energy.

\subsection{Specific model for the scattering rate}

The model scattering rate for a
quasi-one-dimensional system that we consider is
\begin{equation}
\label{tauont2}
{1 \over \tau(k_y)} = {1 \over \tau_0} +
A \sin^2\left({b k_y\over2}\right) \ ,
\end{equation}
where the first term does not vary over
the Fermi surface and $A$ is the strength of the cold spots.
The second term
determines the periodicity of the spots on
the Fermi surface (see Fig.~\ref{coldspot}).
This is a quasi-one-dimensional version of the model
considered by Ioffe and Millis.\cite{ioffe:cs}
If the cold spots are due to superconducting
fluctuations then the superconducting
phase would have nodes in the energy gap at
$(k_x,k_y)=(\pm k_F, 0)$.


Ioffe and Millis took the scattering time $\tau_0$ 
to be the sum of an impurity
part and a temperature dependent part \cite{ioffe:cs}
\begin{equation}
{1 \over \tau_0}={ 1 \over \tau_{imp} }+{T^2 \over T_0} \ ,
\end{equation}
where $T_0$ is an energy scale of the order of the Fermi temperature.

\subsection{Zero and high field limits}

{\it Zero field limit:}
The interlayer conductivity, when $\vec{B}=0$, is given by \cite{AM}
\begin{equation}
\label{cs1}
\sigma_{zz}(B=0)= { e^2\over 4 \pi^3 } \int \tau(\vec{k}(t))
v_z(\vec{k}){v}_z(\vec{k}) d^3\vec{k} \ ,
\end{equation}
where $f(E)$ is the Fermi function, $\tau(\vec{k})$ the
momentum-dependent scattering time and ${v}_z(\vec{k})$ is the
electron velocity perpendicular to the layers. Now in
zero magnetic field the velocities are constant thus
the conductivity becomes
\begin{equation}
\sigma_{zz}(B=0)= { e^2\over 4 \pi^3} \int v_z(k_z)^2
\hspace{1mm}\tau(k_y)
\delta(E_F-\epsilon(\vec{k})) d^3\vec{k} 
= {2 e^2 c t_c^2 \over \pi \hbar^3 b v_F} <\tau> ,
\end{equation}
where $<\tau>$ is the average of the lifetime
of the carriers on the Fermi surface.

{\it High field limit:}
At high fields (as $B \rightarrow \infty $) the term
$\exp\left[ -\int_{0}^{t} {du \over \tau(u)}\right]$ 
in (\ref{tauont})
oscillates rapidly, therefore we replace the scattering rate term
by its average over the Fermi surface, $\left<{1 \over \tau}\right>$,
where $<..>$ denotes the average. Thus we obtain
\begin{equation}
\exp\left[ -\int_{0}^{t} du \left<{1 \over \tau}\right> \right] =
\exp\left[ -t \left<{1 \over \tau}\right> \right]
\end{equation}
and evaluating $\bar{v}_z(\vec k)$(see Eq.~\ref{tauont}) we get
\begin{equation}
\int_{0}^{\infty} dt \exp\left[ -t \left<{1 \over \tau}\right> \right] =
{1 \over \left<{1 \over \tau}\right>} \ ,
\end{equation}
provided that the velocity, $v_z(\vec{k})$, is independent of the time.
The conductivity can then be simplified to
\begin{equation}
\sigma_{zz}(B=\infty) 
= {2 e^2 c t_c^2 \over \pi \hbar^3 b v_F} 
{1 \over \left<{1 \over \tau}\right>} \ .
\label{high}
\end{equation}
Combining the results forboth the high and low field limits
gives
\begin{equation}
{\rho_{zz}(B=\infty) \over \rho_{zz}(B=0) } = < \tau> 
\left<{1 \over \tau}\right>
\label{scat}
\end{equation}
A similar result was obtained by Zheleznyak and Yakovenko\cite{zhel:cs}.

\subsection{Positive Magnetoresistance}
By using the Cauchy-Schwarz
inequality it can be shown that the right hand side of (\ref{scat})
must be greater than or equal to unity. Thus, the saturating value
of the magnetoresistance is always positive. If $f(\vec{k})$ and
$g(\vec{k})$ are functions defined
on the Fermi surface we can define an inner product
\begin{equation}
(f,g) = \int_{FS} d^2k f(\vec{k}) g(\vec{k}) \ ,
\end{equation}
where the integral is over the Fermi surface.
The Cauchy-Schwarz inequality implies that
\begin{equation}
|(f,g)| \leq \| f \|\hspace{2mm} \| g\| \ ,
\end{equation}
where $\| f\|$ denotes the norm of $f$ defined by
$\| f \| = (f,f)^{1/2}$.
We set $f(\vec{k})= \sqrt{{1 \over \tau(\vec{k})}}$
and $g(\vec{k}) = {1 \over f(\vec{k})}$
and square both sides to obtain
\begin{equation}
1 \leq \int_{FS} {1 \over \tau(\vec{k})} d^2k
\hspace{2mm} \int_{FS} \tau(\vec{k}) d^2k =
\left<{1 \over \tau(\vec{k})} \right>< \tau(\vec{k})>
= {\rho_{zz}(B=\infty) \over \rho_{zz}(B=0)} \ .
\end{equation}
This shows that
the resistance at high fields will always be larger
than the resistance at zero field. Note that this result
does not depend on the particular functional form
for the variation of the 
scattering rate over the Fermi surface.
\section{Interlayer conductivity in zero field}
\label{cond-zero}
Substituting the scattering rate (\ref{tauont2}) and
the velocity in the $z$-axis direction, $v_{z} =
{2c t_c \over\hbar}\sin(ck_z)$, into
the conductivity (\ref{cs1}), we obtain
\begin{equation}
\sigma_{zz}(B=0)= { e^2\over 4 \pi^3 \hbar v_F} \left({2c t_c \over\hbar }\right)^2
\int_{-\pi/c}^{\pi/c} \sin(c k_z)^{2} \hspace{2mm}dk_z
\int_{-\pi/b}^{\pi /b}
{ dk_y \over {1 \over \tau_0 }+A \sin\left({b k_y \over 2}\right)^2 } \ .
\end{equation}
Performing the integrals gives
\begin{equation}
\sigma_{zz}(B=0)= {2 e^2 c t_c^2 \tau_0 \over \pi \hbar^3 b v_F}
{1 \over \sqrt{1 + A \tau_0}}
\label{cszero}
\end{equation}
and in the absence of cold spots ($A=0$) we get
\begin{equation}
\sigma_{zz}(A=0)= {2 e^2 c t_c^2 \tau_0 \over \pi \hbar^3 b v_F} \ .
\end{equation}
If $1/\tau_0 \sim T^2$, $A$ is independent of temperature and
$A\tau_0 >> 1$ then 
$\rho_{zz} \sim T$. 
The different temperature dependences that have been
observed in the Bechgaard salts are summarised in Table I.

\section{THE INTERLAYER CONDUCTIVITY IN THE PRESENCE OF A
MAGNETIC FIELD}
\label{cond-all}
\subsection{Magnetic field parallel to the least conducting axis}
\label{condc-axis}

We now show how when the field and current are both
parallel to the c-axis that the cold spots
produce a positive magnetoresistance.
For the dispersion relation (\ref{disper})
the components of the group velocity are
\begin{equation}
{\vec{v}}=\frac{1}{\hbar }\overrightarrow{\nabla }_{k}\epsilon ={1\over \hbar }
\pmatrix { \hbar v_F \cr 2bt_{b}\sin (bk_{y})\cr
2ct_{c}\sin (ck_{z})} \ .
\end{equation}
The rate of change of the wave vector $ \vec{k}(t) $, in
a magnetic field
given by $ \vec{B}=(0,0,B) $, is
\begin{equation}
\label{wavevectcoldspots}
{d\vec{k}\over dt}=-\frac{e}{\hbar^{2} }\overrightarrow{\nabla }_{k}
\epsilon \times \overrightarrow{B}={1\over \hbar }\pmatrix {-2beBt_{b}
\sin (bk_{y})/\hbar\cr e v_F B \cr 0
}\hspace {100pt}\pmatrix {a\cr b\cr c} \ .
\end{equation}
In order to calculate the time dependence of $\vec{k}(t)$ we
integrate Eq.~(\ref{wavevectcoldspots}), giving
\begin{eqnarray}
k_z(t) & = & k_z(0) \\
k_y(t) & = & k_y(0) + {\omega_0 \over b} t \ ,
\end{eqnarray}
where
\begin{equation}
\omega_0 ={e v_F B b \over \hbar}
\end{equation}
is the frequency with which the electron traverses
the Fermi surface. The $z$-component
of the group velocity is then 
\begin{equation}
\label{vzcs}
v_z= {2 c t_c \over \hbar} \sin(c k_z(0)) \ .
\end{equation}
Substituting $k_y(t)$ into (\ref{tauont2}) and evaluating the
exponential in (\ref{tauont}) we obtain
\begin{equation}
\label{vbarcs}
\bar{v}_z(\vec k)= v_z(k_z(0)) \exp\left[{A \over 2 \omega_0} \sin(b k_y(0)) \right]
\int_{-\infty}^{0} dt \exp{\left[{(A\tau_0 + 2) t \over 2 \tau_0}
-{A \over 2 \omega_0} \sin(b k_y(0)+ \omega_0 t)\right] } \ .
\end{equation}
We introduce the modified
Bessel generating function\cite{as} for
\begin{eqnarray}
\label{modBesselGF}
\exp{\left[-{A \over 2 \omega_0} \sin(b k_y(0)+ \omega_0 t)\right] } & = &
I_0\left(-{A \over 2 \omega_0}\right)+ 2 \sum_{k=0}^{\infty}
(-1)^k I_{2 k + 1}\left(-{A \over 2 \omega_0}\right)
\sin((2k+1)(b k_y(0)+ \omega_0 t)) \\ \nonumber
& &
+ 2 \sum_{k=1}^{\infty} (-1)^k I_{2 k}\left(-{A \over 2 \omega_0}\right)
\cos((2k)(b k_y(0)+ \omega_0 t))
\end{eqnarray}
and perform the integral over $t$ to obtain
\begin{eqnarray}
\label{a1}
\bar{v}_z(\vec k) & = & v_z(k_z(0)) \exp\left[
\left(-{A \over 2 \omega_0}\right) \sin(b k_y(0)) \right]
\left\{ {I_0\left(-{A \over 2 \omega_0}\right) \over C} \right. \\ \nonumber
& & + 2 \sum_{k=0}^{\infty}
(-1)^k I_{2 k + 1}\left(-{A \over 2 \omega_0}\right)
\left[ {-(2k+1)\omega_0 \cos(b(2k+1)k_y(0)) + C\sin(b(2k+1)k_y(0))
\over C^2 + (2k+1)^2 \omega^2_0}\right] \\ \nonumber
& & \left. + 2 \sum_{k=1}^{\infty}
(-1)^k I_{2 k}\left(-{A \over 2 \omega_0}\right)
\left[ {(2k)\omega_0 \sin(b(2k)k_y(0)) + C\cos(b(2k)k_y(0))
\over C^2 + (2k)^2 \omega^2_0}\right]\right\} \ ,
\end{eqnarray}
where $C = {A\tau_0 + 2 \over 2 \tau_0}$. A similar substitution
can be made for the $\exp\left[{A \over 2 \omega_0} \sin(b k_y(0)) \right]$
term in (\ref{a1}) by seting $t=0$ in (\ref{modBesselGF}). Multiplying
out all terms, we note that the only terms that survive the integral
over $k_y(0)$ are those whose indicies
in the summations are equal. Performing the integrals in $k_y(0)$ and
$k_z(0)$ the conductivity becomes

\begin{equation}
{ \sigma_{zz}(B) \over \sigma_{zz}(A=0)} = {1 \over (1+{A\tau_0 \over 2})}
\sum_{k = -\infty}^{\infty} {(-1)^k \hspace{3mm}
I_k\left({A \over 2 \omega_{0}}\right)^2\over 1\hspace{2mm}+
\hspace{2mm} {4 k^2 \omega^2_{0} \tau_0^2 \over (2+A\tau_0)^2} } \ .
\label{condcoldspots}
\end{equation}
In the Appendix we present an
alternative form for this expression that is more
stable for numerical evaluation. Fig.~{\ref{p5a} shows
the dependence of the interlayer resistivity
on the strength of the magnetic field
at various values of the parameter $A\tau_0$.

{\it {High field limit:}} The conductivity, as
${A \over \omega_{0}} \rightarrow 0$, is
simplified by the limiting form for small arguments
of the modified Bessel function

\begin{equation}
I_{k}(z) \sim {({1 \over 2 }z)^k \over \Gamma(k+1)}
\hspace{15mm} (k \neq -1,-2,...)
\end{equation}
and so the $k=0$ term dominates (\ref{condcoldspots})
giving
\begin{equation}
\label{highfield}
{ \rho_{zz}(\omega_{0} \gg A) \over \rho_{zz}(A=0)} =1+{A\tau_0 \over 2} \ .
\end{equation}
This agrees with the general result (\ref{high}).

\subsection{Magnetic field parallel to the $b$-axis}
Chashechkina and Chaikin found
that for (TMTSF)$_2$ClO$_4$ under 6 kbar pressure\cite{chas:cs},
the interlayer resistivity (for field directed along the b-axis)
deviates from the quadratic field dependence which is
predicted from simple Boltzmann transport theory.
Although it is quadratic at low fields the resistivity
becomes approximately linear at higher fields.
Kohler's rule is obeyed.
In contrast,
for (TMTSF)$_2$PF$_6$ at 10 kbar the interlayer
resistivity saturates above fields of about 2 tesla\cite{chas2:cs,lee:cs}.

With a magnetic field given by $\vec{B}=(0,B,0)$ the rate of change
of the wave vector is $d\vec{k}/dt =
({2eBct_c\sin(ck_z) \over \hbar^2 },0,-{ev_F B \over \hbar })$. From this
the $z$-axis velocity is calculated to be
$v_z(k_z)= {2ct_c \over \hbar} \sin(ck_z(0)-\omega_{0c}t)$, where
$\omega_{0c}= c\omega_0/b$. In this case, when the magnetic
field is parallel to the $b$-axis, $k_y$ is constant
and so $\tau$ is
not a function of time. Thus the electron trajectories are either
in or out of the cold spot region, but never swept through them. One
can write Eq.~(\ref{tauont}) as
\begin{equation}
\bar{v}_z(\vec k)=
\int_{-\infty}^{0} dt\hspace{5pt} v_{z}(\vec{k}(t))
\exp\left[ - {t \over \tau(k_y)}\right] \ .
\end{equation}
After the appropriate substitution for the scattering rate
and $z$-axis velocity we obtain
\begin{equation}
\bar{v}_z(\vec k)= {2ct_c \over \hbar}\left[
{ \omega_{0} \cos(ck_z(0)) -R \sin(ck_z(0))\over R^2 +\omega_{0}^2 }\right] \ ,
\end{equation}
where $R={1 \over \tau_0}+A\sin\left( {bk_y(0) \over 2}\right)^2$
and for simplicity here we set $b=c$ so $\omega_{0c}=\omega_0$. 
The conductivity can then be written as
\begin{equation}
\sigma_{zz}(B)= {e^2 \over 4 \pi^3 \hbar v_F b}
\left({2ct_c \over\hbar }\right)^2
\int_{0}^{2\pi/b} dk_y(0) \int_{0}^{2\pi/c} dk_z(0) \sin(ck_z(0))
\left[{ \omega_{0} \cos(ck_z(0)) -R \sin(ck_z(0))\over R^2 +\omega_{0}^2 }\right] \ .
\end{equation}
Performing the integral over $dk_z(0)$ we obtain
\begin{equation}
\sigma_{zz}(B)= {e^2 \over 4 \pi^2 c \hbar v_F b} \left({2ct_c \over\hbar }\right)^2
\int_{0}^{2\pi/b} dk_y(0)
{\left( {1 \over \tau_0}+A\sin\left( {bk_y(0) \over 2}\right)^2\right)
\over
\left( {1 \over \tau_0}+A\sin\left( {bk_y(0) \over 2}\right)^2\right) ^2
+ \omega_0^2}
\end{equation}
and integrate to give

\begin{equation}
\label{condzz-B-baxis}
{\sigma_{zz}(B) \over \sigma_{zz}(A=0)}=
{\sin\left({\arctan\left({1\over \omega_0\tau_0}\right)+
\arctan\left({1+A\tau_0\over \omega_0\tau_0}\right) \over 2}\right)
\over [1+ ( \omega_0\tau_0)^2]^{1/4}
\hspace{2mm}[(1+A\tau_0)^2+ (\omega_0\tau_0)^2]^{1/4}} \ .
\end{equation}

{\it High field limit:} If $\omega_0 \gg 1/\tau_0$ and $\omega_0 \gg A$
we can expand in ${1 \over \omega_0\tau_0}$ to second order to
obtain
\begin{equation}
{\sigma_{zz}(B) \over \sigma_{zz}(A=0)}=
{2+A\tau_0 \over2 }{1 \over (\omega_0\tau_0)^2 }.\
\end{equation}
Thus, at high fields the resistivity is quadratic in field
and does not saturate. This is inconsistent with
the experimental results on TMTSF$_2$X cited above.

{\it Low field limit:} Here we expand in $\omega_0\tau_0$ to second order to obtain
\begin{equation}
{\sigma_{zz}(\omega_0\tau_0 \ll 1) \over \sigma_{zz}(A=0)}= {1 \over \sqrt{1+A\tau_0}}-
{ (8+A\tau_0(8+3A\tau_0))\over 8(1+A\tau_0)^{5/2} }(\omega_0\tau_0)^2 \ ,
\end{equation}
where we can write, after simplifying, the resistivity as
\begin{equation}
\label{quadfit}
{\rho_{zz}(B) \over \rho_{zz}(A=0)}= \sqrt{1+A\tau_0}\left(
1+ {(8+A\tau_0(8+3A\tau_0)) (\omega_0\tau_0)^2 \over 8(1+A\tau_0)^2}\right)
\end{equation}
A comparison of the result (\ref{condzz-B-baxis}) with the
quadratic form (\ref{quadfit}) is shown in Fig.~\ref{p5ex-qd},
for $A\tau_0 =1,10$. In the absence of cold spots
Boltzmann transport theory predicts
a quadratic field dependence for all fields.
The plot shows that the quadratic fit (dashed line)
deviates form the exact solution (solid line) at large fields.
As the strength of the cold spots increase the deviations increase further,
while the exact solution becomes increasingly linear at small fileds.
Note that the low-field quadratic fit always lies above
the actual result, as is observed in
(TMTSF)$_2$ClO$_4$ at 6 kbar.\cite{chas:cs}.

{\it Kohler's rule: } 
Equation (\ref{condzz-B-baxis})
shows that the resistance depends on three
parameters: $\omega_0 $ which is linearly
proportional to the magnetic field, 
the scattering time $\tau_0$ 
and $A$ the parameter that determines the strength of the cold spots.
Since $\tau_0$ and $A$ can both depend on temperature,
we can analyse the temperature
and field dependence of the magnetoresistance in terms of
Kohler's rule\cite{pipp}. Kohler's rule is known to hold when
there is a single species of charge carrier and the
scattering time $\tau$
is the same at all points on the Fermi surface\cite{mcken}.
The dependence of the resistivity on the field in (\ref{condzz-B-baxis})
is contained
in the quantity $\omega_0\tau_0$ and the temperature dependence
of $A\tau_0$. In zero field the conductivity is given
by (\ref{cszero}).
The field dependence
of the magnetoresistance, with different scattering times, can be related
by scaling the field by the zero-field resistivity $\rho_{zz}(B=0, A\tau_0)$.
To obtain a Kohler's plot we plot
${\rho_{zz}(B, A\tau_0)\over \rho_{zz}(B=0, A\tau_0)}$ versus
${B \over \rho_{zz}(B=0, A\tau_0)}$.
In order to do this we re-arrange
Eq.~(\ref{condzz-B-baxis}) to give
\begin{equation}
{\sigma_{zz}(B,A\tau_0) \over \sigma_{zz}(B=0,A\tau_0)}=
{\sin\left({\arctan\left({1\over \omega_0\tau_0}\right)+
\arctan\left({1+A\tau_0\over \omega_0\tau_0}\right) \over 2}\right)
\over [1+ ( \omega_0\tau_0)^2]^{1/4}
\hspace{2mm} \left[1+ {(\omega_0\tau_0)^2
\over (1+A\tau_0)^2}\right]^{1/4}} \ ,
\end{equation}
and plot the inverse of this against $\omega_0\tau_0 / \sqrt{1+A\tau_0}$,
because $\rho_{zz}(B=0, A\tau_0) \propto \sqrt{1+A\tau_0} /\tau_0$.
Fig.~\ref{p5kohl}
shows such a plot for various values of $A\tau_0$. The figure shows that Kohler's
rule is violated at high fields and for $A\tau_0 \stackrel{>}{\sim} 5$;
if it held all the curves would collapse onto a single curve.

\subsection{Magnetic field in the $b-c$ plane}
For rotations of the magnetic field in the $b-c$ plane
experiments on (TMTSF)$_2$ClO$_4$
at ambient\cite{dan:cs} and 6 kbar\cite{chas:cs} pressure
and 
(TMTSF)$_2$PF$_6$ at 6 kbar\cite{lee:cs}
find that the angular dependence
of the interlayer magnetoresistance
has dips at the magic angles superimposed on
roughly the angular dependence predicted by
semi-classical transport theory.
The magnetoresistance is minimum when the magnetic
field and the current are parallel
and a maximum when the field and current are perpendicular.
This is in contrast to the anomalous behaviour seen in
(TMTSF)$_2$PF$_6$ at 10 kbar \cite{clarke,chas2:cs}
where the opposite is observed: the background magnetoresistance
only depends on the component of the field
perpendicular to the layers, 
$\rho_{zz} \sim (B \cos \theta)^{1.3}$.

Following a similar procedure as in Section \ref{condc-axis} the
rate of change of the wave vector, in
a magnetic field
given by $ \vec{B}=(0,B\sin\theta,B\cos\theta) $, is
\begin{equation}
{d\vec{k}\over dt}=-\frac{e}{\hbar^{2} }\overrightarrow{\nabla }_{k}
\epsilon \times \overrightarrow{B}={1\over \hbar^2 }\pmatrix {-2beBt_{b}\cos\theta
\sin (bk_{y}) \cr e v_F \hbar B\cos\theta \cr -e v_F \hbar B\sin\theta
}\hspace {100pt}\pmatrix {a\cr b\cr c} \ .
\end{equation}
The velocity in the $c$-direction ($z$-axis) can then be written
\begin{equation}
v_z(k_z)={2 c t_c \over \hbar} \sin(ck_z(0) -\omega_c t)
\end{equation}
and $\bar{v}_z(\vec k)$, from Eq.~(\ref{tauont}), can be calcualted
by making the appropriate substitutions for the scattering
rate and the $c$-axis velocity, giving
\begin{equation}
\bar{v}_z(\vec k)= {2 c t_c \over \hbar}
\exp\left[{A \over 2 \omega_B} \sin(b k_y(0)) \right]
\int_{-\infty}^{0} dt \sin(ck_z(0) -\omega_c t)
\exp{\left[{(A\tau_0 + 2) t \over 2 \tau_0}
-{A \over 2 \omega_B} \sin(b k_y(0)+ \omega_B t)\right] } \ ,
\end{equation}
where $\omega_B = {e b B v_F \cos\theta \over \hbar}= \omega_{0}\cos\theta $ and
$\omega_c = {e c B v_F \sin\theta \over \hbar}$.
Substitution of the appropriate modified Bessel generating functions
and performing the integral over $t$ gives
\begin{eqnarray}
\bar{v}_z(\vec k) & = & {2 c t_c \over \hbar}
\exp\left[
\left(-{A \over 2 \omega_B}\right) \sin(b k_y(0)) \right]
\left\{ I_0\left(-{A \over 2 \omega_B}\right)
\left[ 
{\omega_c \cos(c k_z(0))+ C \sin(c k_z(0)) \over C^2 
+ \omega_c^2}\right] \right. \\ \nonumber
& & + \sum_{k=0}^{\infty} 
(-1)^k I_{2 k + 1}\left(-{A \over 2 \omega_B}\right) 
\left[ {C \cos(ck_z(0) -(2k+1)bk_y(0))-
((2k + 1)\omega_B+\omega_c)\sin(ck_z(0) -(2k+1)bk_y(0))
\over {C^2 + ((2k+1)\omega_B+\omega_c)^2}} \right. \\ \nonumber
& & \hspace{70pt} -\left. {C \cos(ck_z(0) +(2k+1)bk_y(0))-
((2k + 1)\omega_B-\omega_c)\sin(ck_z(0) +(2k+1)bk_y(0))
\over {C^2 + (-(2k+1)\omega_B+\omega_c)^2}}
\right] \\ \nonumber
& & + \sum_{k=1}^{\infty} 
(-1)^k I_{2 k }\left(-{A \over 2 \omega_B}\right) 
\left[ {C \sin(ck_z(0) -(2k)bk_y(0))+
((2k )\omega_B-\omega_c)\cos(ck_z(0) -(2k)bk_y(0))
\over {C^2 + (2k\omega_B+\omega_c)^2}} \right. \\ \nonumber
& & \hspace{70pt} +\left. {C \sin(ck_z(0) +(2k)bk_y(0))+
((2k)\omega_B+\omega_c)\cos(ck_z(0) +(2k)bk_y(0))
\over {C^2 + (-(2k)\omega_B+\omega_c)^2}}
\right] \ ,
\end{eqnarray}
where $C ={1 \over \tau_0}+{A \over 2}$. Performing the 
integrals over $k_y(0)$ and $k_z(0)$ one obtains
\begin{equation}
{\sigma_{zz}(B) \over \sigma_{zz}(A=0)}= \left({1 \over1+ {A\tau_0 \over 2} }\right)
\sum_{k=-\infty}^{\infty} {(-1)^k I_k({A \over 2 \omega_B})^2
\over 1 + { 4 \tau_0^2 (k \omega_B + \omega_c)^2\over (2 + A \tau_0)^2} } \ .
\label{sigzz-bc}
\end{equation}
Based on this expression we
expect to see features in the angular dependence when
\begin{equation}
k={\omega_c \over \omega_B}= {c \over b}\tan\theta \ .
\label{MA_condition}
\end{equation}
Due to the alternating sign in the summation,
when the index $k$ is even one expects to see dips,
while when $k$ is odd one expects peaks in the resistivity.
A plot of the interlayer resistivity versus the field tilt angle $\theta$
is shown in Fig.~\ref{p5-sigbc} for several parameter values. 
It can be seen that only the $k=1$ resonance is noticeable,
and only for very large
fields ($\omega_0 \tau_ 0 > 100$).
Experimentally the magic angle effects are seen at much
lower fields. Furthermore, one always sees dips and not
peaks at the magic angles.
\section{Conductivity parallel to the chains}
\label{condx-axis}
Measurements of the resistivity parallel to the $a$ axis
for rotations of the magnetic field in the $b-c$ plane
show similar behavior as for the interlayer resistivity
\cite{dan:cs,chas:cs,clarke,chas2:cs}.
Magic angle effects are superimposed on a background 
magnetoresistance which has a semiclassical angular
dependence for 
(TMTSF)$_2$ClO$_4$ and is anomalous
for (TMTSF)$_2$PF$_6$ at 10 kbar.
For the latter a
power law field dependence of the $a$-axis resistivity was
found with the field in the $c$-axis
direction by Kriza {\it et al.} \cite{kri:cs},
$\rho_{xx}(B) - \rho_{xx}(0) \propto B^{3/2}$.

The conductivity parallel to the chains ($\sigma_{xx} $) is
calculated in a similar manner
to the interlayer conductivity, where the magnetic field is rotated
in the $b-c$ plane. Calculating
the velocity in the $x$-axis direction ($v_x = v_F$), we can substitute
this and our specific model for the
scattering rate into Eq.~(\ref{condcs}) and (\ref{tauont}) to obtain

\begin{eqnarray}
\sigma_{xx} & = & {e^2 \over 4 \pi^3 \hbar v_F}
\int_{0}^{2\pi/c} dk_z(0)
\int_{0}^{2\pi/b} dk_y(0) v_F \exp\left[{A \over 2
\omega_B} \sin(b k_y(0)) \right]
\nonumber \\
& & \times \int_{-\infty}^{0} dt\hspace{5pt} v_F
\exp{\left[{(A\tau_0 + 2) t \over 2 \tau_0}
-{A \over 2 \omega_B} \sin(b k_y(0)+ \omega_B t)\right] } \ .
\end{eqnarray}
Performing the integral and simplifying we obtain 
\begin{equation}
{\sigma_{xx} \over \sigma_{xx}(A=0)}= \left({1 \over1+ {A\tau_0 \over 2} }\right)
\sum_{k=-\infty}^{\infty} {(-1)^k I_k({A \over 2 \omega_0 \cos\theta})^2
\over 1 + { 4 (k \tau_0 \omega_0 \cos\theta)^2\over (2 + A \tau_0)^2} } \ .
\end{equation}
Note that for $\theta =0$ this will give the same field
dependence for the conductivity in the least conducting
direction (compare Eq.~{\ref{condcoldspots} and Fig.~\ref{p5a}).

The angular dependence of the resistivity $\rho_{xx}
\equiv 1/\sigma_{xx}$ given by the equation above is plotted in
Fig.~\ref{p5xx} for two values of $A\tau_0$ and
$\omega_{0}\tau_0$. We see that some similarities exist
between theory and experimental results on
(TMTSF)$_2$PF$_6$ at 9.5 kbar pressure for rotations
of the magnetic field in the $b-c$ plane \cite{chas2:cs,dan2:cs}, in that the
resistivity is large for magnetic field angles close to
$\theta$=$0^o$ and decreases as $\theta$ approaches $90^o$.
Furthermore, the resistivity only depends on the
component of field perpendicular to the layers; that is,
$\omega_B =\omega_0 \cos\theta$. We tried fitting the
field dependence to a power law of the form
$\rho_{xx} \sim (B\cos\theta)^\alpha$ but found this
only applied over very limited field ranges.
The calculated angular dependence of $\rho_{xx}$ also
differs from the observed angular dependence in that
no magic angle features are present in the calculated
$\rho_{xx}(\theta)$.
\section{Conclusion}

We have considered a modification of standard
Fermi liquid and Boltzmann transport theory in
which there are large variations of the quasiparticle
scattering rate over a quasi-one-dimensional
Fermi surface. The goal was to see to what
extent such a model could explain the anomalous
properties of the magnetoresistance of
the quasi-one-dimensional organic metals,
(TMTSF)$_2$X.
Table \ref{table1:cs} 
gives a brief comparison of the results of
our calculations for a cold spots model
with experimental results.
Although the model can explain a number of unusual
features such as having a large
magnetoresistance when the field and current
are parallel there are several important 
discrepancies.  Although the model does
give magic angle effects they are orders of magnitude 
smaller than is observed experimentally.  
In particular explaining the origin of the 
magic angle effect and
why in 
(TMTSF)$_2$PF$_6$ at 10 kbar
the interlayer resistivity becomes independent of field for
fields parallel to the $b$ axis remains
a considerable challenge.

\acknowledgements
We thank L. Ioffe and J. Merino for helpful
discussions. This work was supported by the Australian Research
Council.

\appendix
\section{ Alternative expression for conductivity}

We now derive an alternative 
expression for Eq.~(\ref{condcoldspots}) 
which is more stable for numerical evaluation.
One can re-write the conductivity in (\ref{condcs}), using
(\ref{vzcs}) and (\ref{vbarcs}), as

\begin{eqnarray}
\sigma_{zz} & = & { e^2\over 4\pi^3 \hbar v_F} \left({2ct_c \over \hbar}\right)^2
\int_{-\pi / c}^{\pi / c} dk_z(0) \sin(c k_z(0))
\int_{0}^{2\pi / b} dk_y(0)\exp\left[ {A \over 2 \omega_{0}} \sin(b k_y(0))\right]
\\ \nonumber
& & \hspace{5mm} \times \int_{-\infty}^{0} dt \sin(c k_z(0))
\exp{\left[{(A\tau_0 + 2) t \over 2 \tau_0}
-{A \over 2 \omega_{0}} \sin(b k_y(0)+ \omega_{0} t)\right] } \ .
\end{eqnarray}
Since $\sin(bk_y(0)+\omega_0 t)$ is periodic in $t$
we can divide the range of integration into segments of length
${2 \pi \over \omega_0}$ and sum the resulting geometric series giving

\begin{eqnarray}
\sigma_{zz} & = & { c e^2\over 4\pi^2 \hbar v_F} \left({2t_c \over \hbar}\right)^2
{1 \over1- \exp\left(-{\pi(A\tau_0+2) \over \omega_{0} \tau_0}\right)}
\int_{0}^{2\pi/b} dk_y(0)\exp\left[ {A \over 2 \omega_{0}} \sin(b k_y(0))\right]
\\ \nonumber
& & \hspace{5mm }\times \int_{0}^{2 \pi } {d\phi \over \omega_{0}}
\exp{\left[-{(A\tau_0 + 2)\over 2 \tau_0}{\phi \over \omega_{0}}
-{A \over 2 \omega_{0}} \sin(b k_y(0)+ b \phi)\right] } \ ,
\end{eqnarray}
where $\phi = \omega_{0} t$. Shifting the integration over $k_y(0)$ by
$-\phi/2$ and re-arranging terms we obtain

\begin{eqnarray}
\sigma_{zz} & = & { c e^2\over 4\pi^2 \hbar v_F} \left({2t_c \over \hbar}\right)^2
{1 \over1- \exp\left(-{\pi(A\tau_0+2) \over \omega_{0}\tau_0}\right)}
\int_{0}^{2\pi/b} dk_y(0)\exp\left[ {A \over 2\omega_{0}}
(\sin(b k_y(0)-\phi/2)-\sin(b k_y(0)+\phi/2))\right]
\\ \nonumber
& & \hspace{5mm} \times \int_{0}^{2 \pi} {d\phi \over \omega_{0}}
\exp{\left[-{(A\tau_0 + 2)\phi\over 2 \omega_{0}\tau_0} \right]}
\end{eqnarray}
which upon simplification and performing the integration over $k_y(0)$ gives

\begin{equation}
{\sigma_{zz}(B) \over \sigma_{zz}(0)} =\left({1 \over \omega_{0}\tau_0}\right)
{1 \over1- \exp\left(-{\pi(A\tau_0+2) \over \omega_{0}\tau_0}\right)}
\int_{0}^{2 \pi} d\phi \hspace{3mm}I_0 \left({A\tau_0 \over \omega_{0}\tau_0} \sin(\phi/2)\right)
\exp{\left[-{(A\tau_0 + 2)\phi\over 2 \omega_{0}\tau_0} \right]} \ .
\end{equation}



\begin{figure}
\centerline{\epsfxsize=14cm \epsfbox{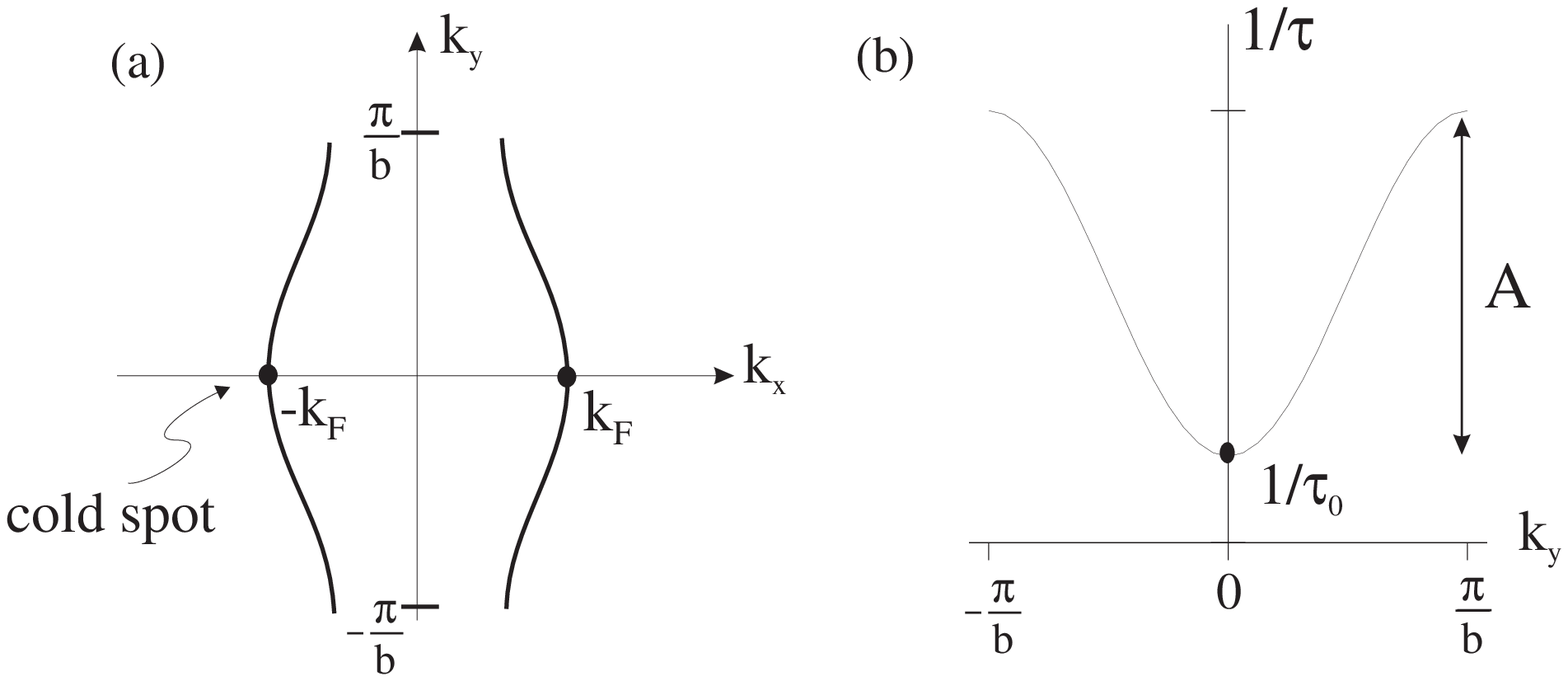}}
\caption{(a) Cold spots on the intra-layer Fermi surface
in a quasi-one-dimensional metal. For a three dimensional
Fermi surface the cold spots become cold strips. A magnetic field
perpendicular to the layers causes electrons
on the Fermi surface to be swept in and out of the cold spots.
(b) Variation of the scattering rate across the Fermi surface.
The strength of the scattering rate at the cold spot is
${1 \over \tau_0}$ and increases by $A$ at the edges of the
Brillouin zone.
\label{coldspot}}
\end{figure}
\begin{figure}
\centerline{\epsfxsize=10cm \epsfbox{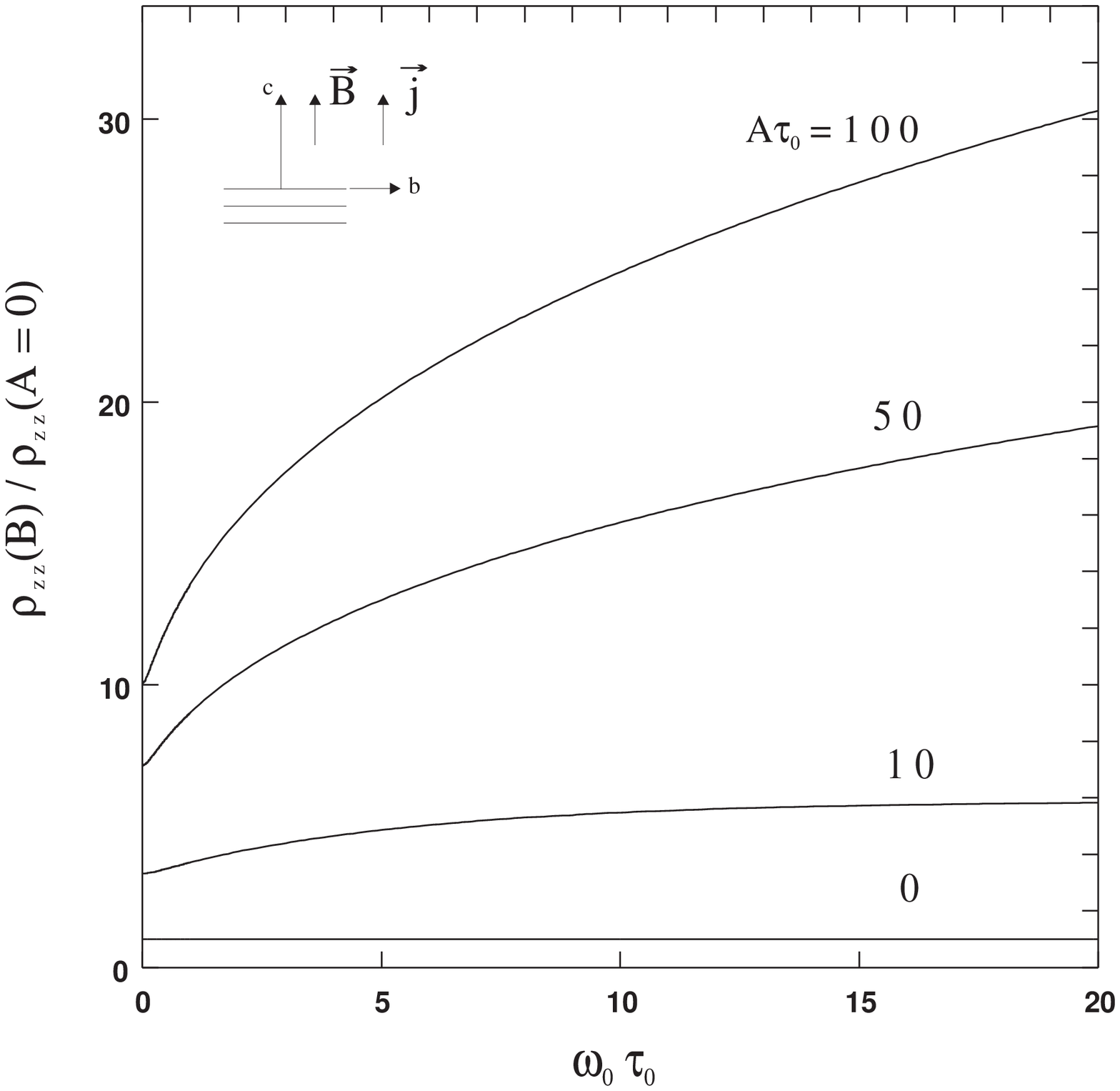}}
\caption{ Dependence of interlayer resistivity on
the strength of the magnetic field
at various values of the parameter $A\tau_0$, which
is a measure of the strength of the scattering
cold spots.
The magnetic field is perpendicular to the
layers and parallel to the current direction and the
$c$-axis (see inset).
In the absence of cold spots ($A=0$) the resistivity is
independent of the field.   As the
strength of the cold spots increases
the magnetoresistance increases and is positive and
non-zero.
For high magnetic fields ($\omega_{0} \gg A$) the
resistivity saturates to
a value given by Eq.~(\ref{highfield}).
\label{p5a}}
\end{figure}
\begin{figure}
\centerline{\epsfxsize=10cm \epsfbox{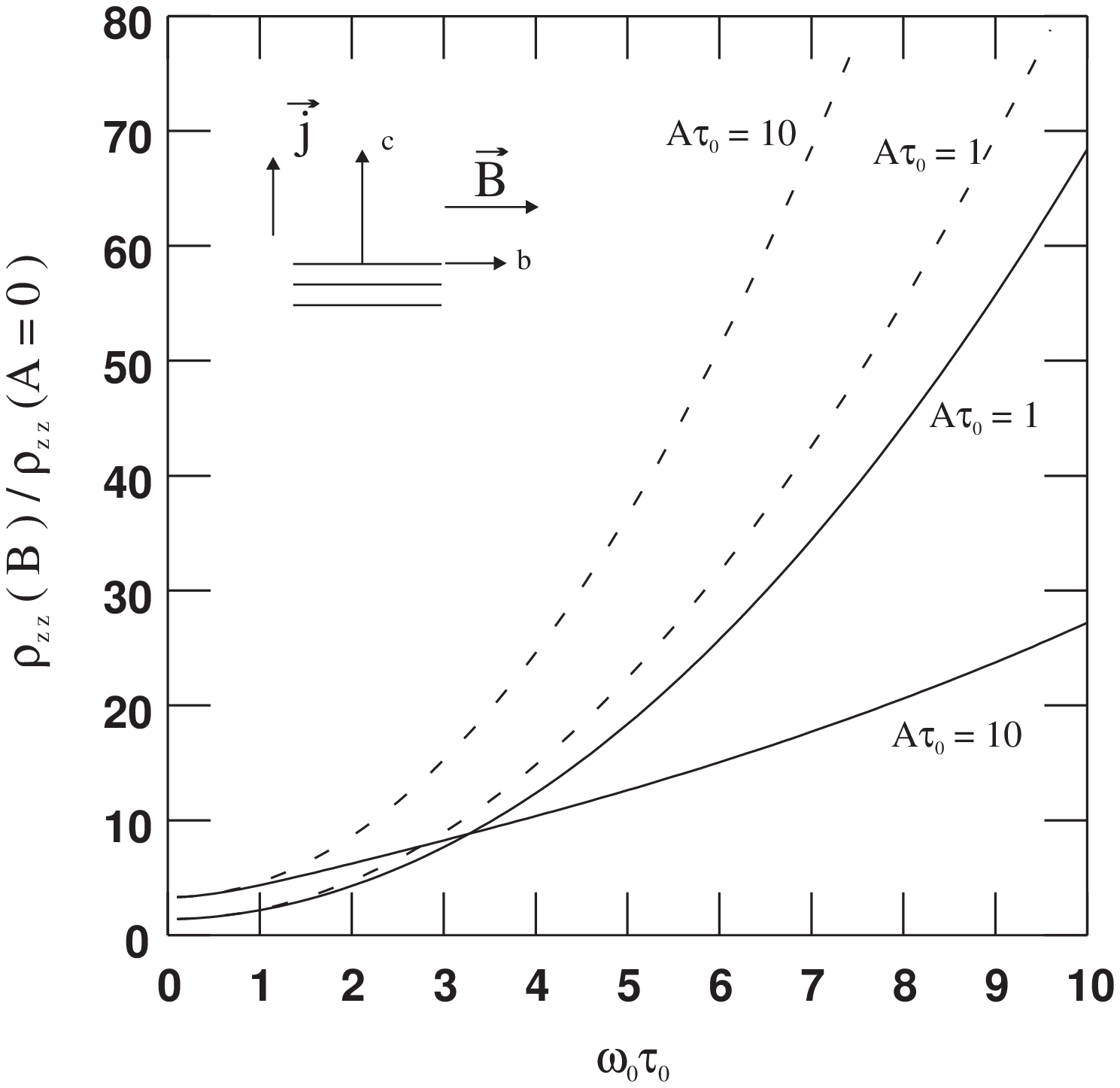}}
\caption{ Field dependence of the interlayer magnetoresistance
when the magnetic field is parallel to the $b$-axis.  
The exact solution (solid line) and the 
quadratic fit to the low field magnetoresistance (dashed line), 
are compared.  
The quadratic form, as predicted from a simple
Boltzmann model \cite{chas:cs} does not fit the exact form at high
fields. This can be compared to experimental results on
(TMTSF)$_2$ClO$_4$ at 6 kbar\cite{chas:cs}. Deviations
from the quadratic form arise due to the variation of the
scattering rate over the Fermi surface.
As the strength of the cold spots increases the
deviation of the low-field fit from the exact solution
increases and the exact solution
becomes increasingly linear at small fields.
Also note that
the quadratic form lies above the exact solution at all values
of $A\tau_0$.
\label{p5ex-qd}}
\end{figure}

\begin{figure}
\centerline{\epsfxsize=10cm \epsfbox{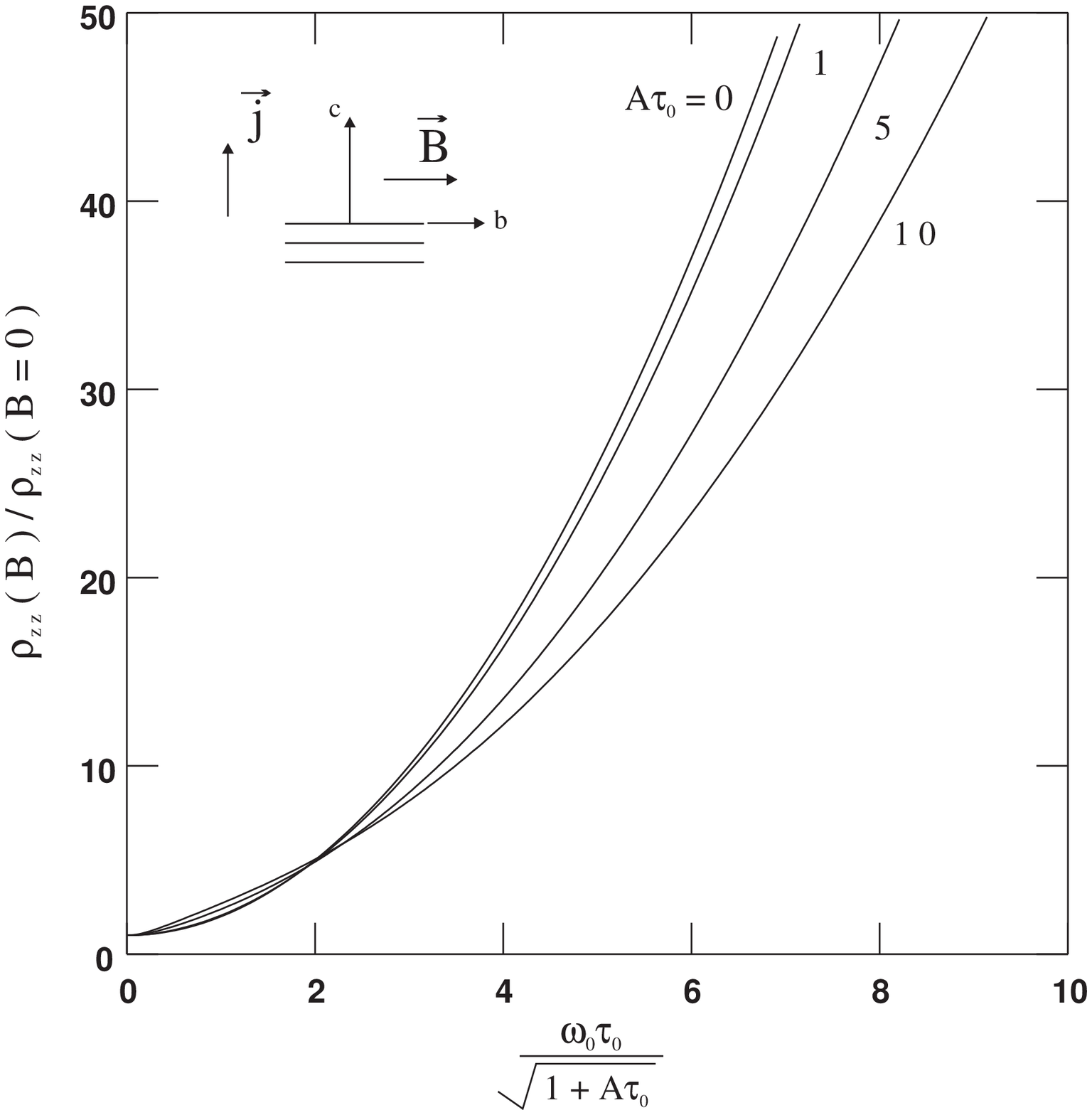}}
\caption{ Kohler's plot of the interlayer magnetoresistance
when the magnetic field is parallel to the $b$-axis.
Plots are shown
for various values of $A\tau_0$, a quantity that can depend on temperature.
The horizontal
axis is proportional to ${B \over \rho_{zz}(B=0)}$.
We see that Kohler's rule is violated since all the curves
do not lie on top of each other. However, the violations are only
significant for large magnetic fields and if the
cold spots are sufficiently strong that $A\tau_0 \geq 5$.
\label{p5kohl}}
\end{figure}
\begin{figure}
\centerline{\epsfxsize=10cm \epsfbox{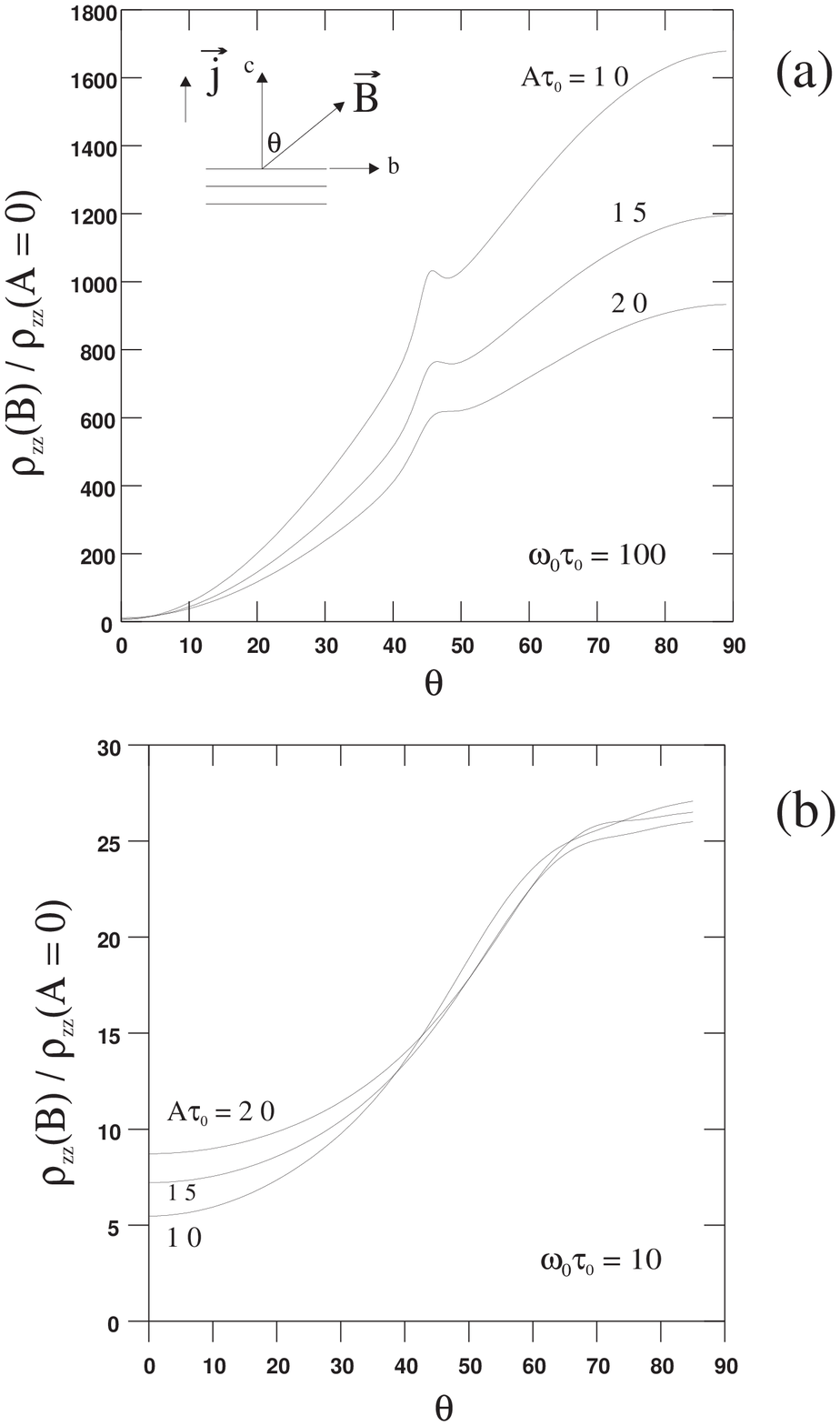}}
\caption{Absence of magic angle effects in the angular dependence of
the interlayer magnetoresistance. The
dependence of the interlayer resistivity on
the magnetic field direction (rotated in the $b-c$ plane)
is shown for various values of $A\tau_0$ and two
values of $\omega_0 \tau_0$ which is 
proportional to the strength of the magnetic field.  
$\theta$ is the angle between
the most conducting direction ($c$-axis) and the
magnetic field (see inset of $(a)$). In contrast
to experimental results on the quasi-one
dimensional metals (TMTSF)$_2$X
one sees a peak
rather than a dip, at $\tan\theta = {b \over c}$. Furthermore,
features at higher order magic angles ($\tan\theta = {n b \over c}$
where $n=2,3, ...$)
are too small to be visible.
\label{p5-sigbc}}
\end{figure}
\begin{figure}
\centerline{\epsfxsize=10cm \epsfbox{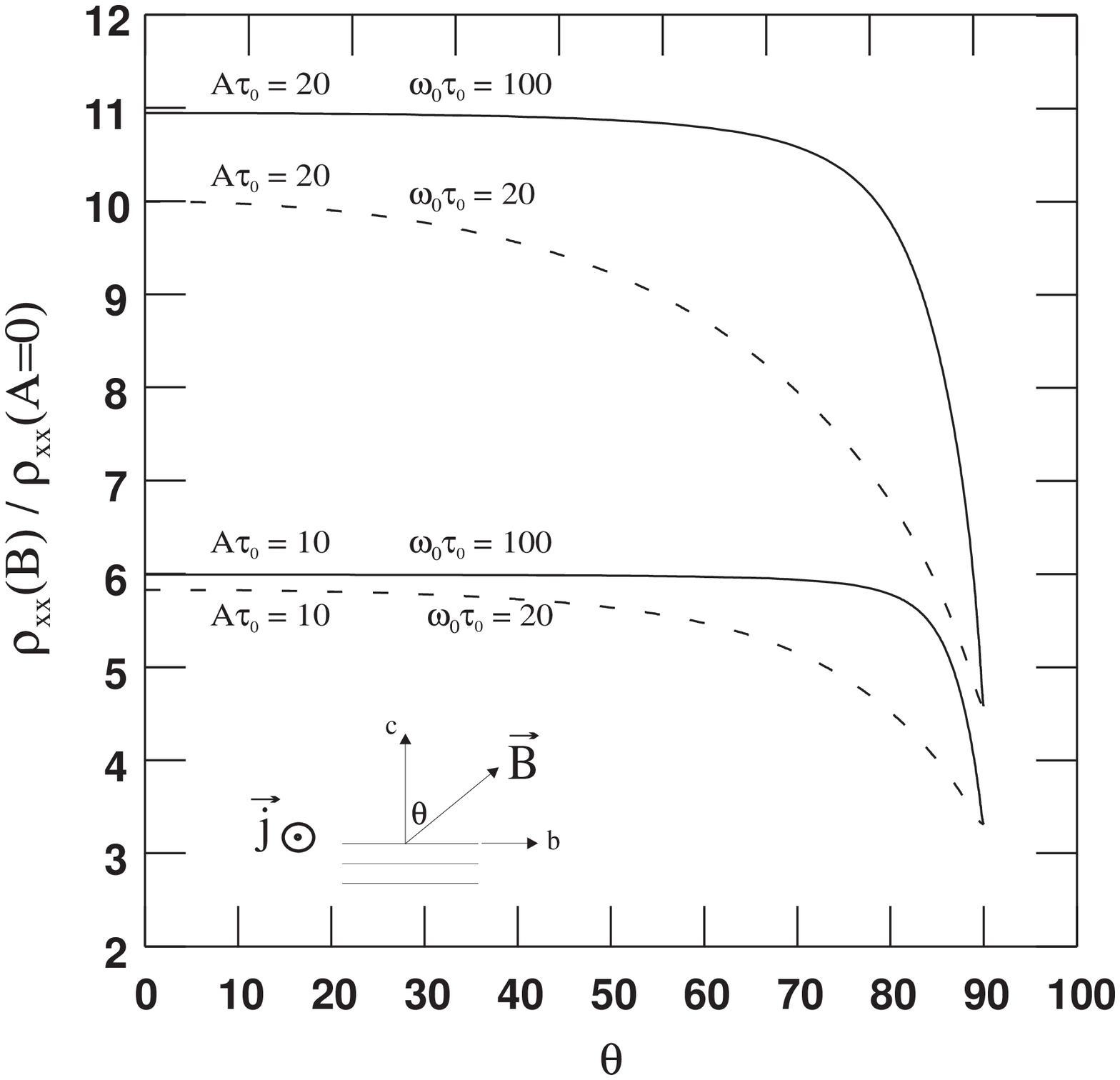}}
\caption{ Angular dependence of the $x$-axis (most
conducting direction) resistivity
on the direction of the magnetic field in the
$b-c$ plane. In comparison to experimental data
on (TMTSF)$_2$PF$_6$ at 9.5 kbar pressure\cite{chas2:cs,dan2:cs},
we see a similarity in
that the interlayer resistance only depends on the
component of field parallel to the $c$ axis and
decreases with increasing angle.
However, no features are present at the magic angles.
\label{p5xx}}
\end{figure}
\newpage
\onecolumn
\widetext
\begin{table}
\caption{Comparison of the observed properties
of the magnetoresistance of (TMTSF)$_2$X  with the theoretical
cold spot model.}
\begin{tabular}{p{1.5in}|ccc}
Effect &X=ClO$_4$(ambient) &X=PF$_6$(9-11 kbar) & Cold spot model \\
\tableline
Magic angle effect in $\rho_{xx}$ & yes & yes & no \\ \hline
Magic angle effect in $\rho_{zz}$ & yes & yes & yes: but too weak \\ \hline
Peaks rather than dips for odd integers & no  & no & yes \\ \hline
Background magnetoresistance only depends on
$\cos\theta$ & no & yes & yes:$\rho_{xx}$, no:$\rho_{zz}$ \\ \hline
Violations of Kohler's rule & no & yes & yes \\
\end{tabular}
\label{table1:cs}
\end{table}
\begin{table}
\caption{The temperature dependence of
the zero-field resistivity of (TMTSF)$_2$X
at various pressures.  We also show if the classical angular
dependence curve is observed in the
particular materials.}
\begin{tabular}{lllll}
X & pressure
& Classical angular dependence & $\rho_{zz}(B=0,T)$& $\rho_{xx}(B=0,T)$ \\
\tableline
ClO$_4$ & ambient& Yes\protect[\onlinecite{dan:cs}] & -- & $T^2$\protect[\onlinecite{kor:cs}] \\
ClO$_4$& 6 kbar& Yes\protect[\onlinecite{chas:cs}] & -- & -- \\
PF$_6$& ambient& -- & $T^2$\protect[\onlinecite{moser:cs}] &
$T^{1.8}$\protect[\onlinecite{moser:cs}],
$T^2$\protect[\onlinecite{moser2:cs}],$T^{1.5}$\protect[\onlinecite{mih:cs}] \\
PF$_6$& 6 kbar& Yes\protect[\onlinecite{lee:cs}] & -- & -- \\
PF$_6$& 8-11 kbar& No\protect[\onlinecite{dan:cs,clarke,chas2:cs}] &
$T^2$\protect[\onlinecite{moser:cs,gor:cs}],
$T$\protect[\onlinecite{clarke}] &
$T^{1.8}$\protect[\onlinecite{moser:cs}] \\
\end{tabular}
\label{table2:cs}
\end{table}

\end{document}